\documentclass[11pt]{IEEEtran}
\usepackage{ownstyle}
\usepackage{color}
\usepackage{hyperref}
\usepackage{bookmark}

\makeatletter
\newlength \figwidth
\if@twocolumn
\else
\fi
\makeatother

\newcommand\doublesum[2]{\sum_{\genfrac{}{}{0pt}{2}{#1}{#2}}}

\title{Feasibility of interference alignment for the\\ MIMO interference channel}
\author{\thanks{A portion of this work was presented at the Information Theory Workshop (ITW) \cite{BCT11a} and the Allerton conference \cite{BCT11}, both in 2011.

G.B. and D.T. have been supported  by the Center for Science of
Information (CSoI), and NSF Science and Technology Center, under grant agreement
CCF-0939370.
D.C. began work on this project while he was a postdoc at the
Mittag-Leffler institute during the program on ``Algebraic geometry with a view
towards applications.'' Since then, he has been supported by NSF grant
DMS-1103856.}
Guy Bresler\thanks{G.B. is currently at LIDS, Dept. of EECS, MIT. Most of the work
in this paper was done when he was at UC Berkeley. (email: gbresler@mit.edu)},
Dustin Cartwright\thanks{D.C. is currently at the Dept. of Mathematics, Yale.
(email: dustin.cartwright@yale.edu)},
David Tse, ̃\IEEEmembership{ Fellow, ̃IEEE}\thanks{D.T. is at Wireless Foundations, Dept. of EECS, UC Berkeley.
(email: dtse@eecs.berkeley.edu)}}

\begin{document}
\maketitle
\begin{abstract}
We study vector space interference alignment for the MIMO interference channel with no time or frequency diversity, and no symbol extensions.
We prove both necessary and sufficient conditions for alignment. In particular,
we characterize the feasibility of alignment for the symmetric three-user channel where all users transmit along $d$ dimensions, all transmitters have $M$ antennas and all receivers have $N$ antennas, as well as feasibility of alignment for the fully symmetric ($M=N$) channel with an arbitrary number of users.

An implication of our results is that the total degrees of freedom available in a $K$-user interference channel, using only spatial diversity from the multiple antennas, is at most~$2$. This is in sharp contrast to the $\frac K2$ degrees of freedom shown to be possible by Cadambe and Jafar with arbitrarily large time or frequency diversity.

Moving beyond the question of feasibility, we additionally discuss computation of the number of solutions using Schubert calculus in cases where there are a finite number of solutions. 
\end{abstract}


\section{Introduction}
\label{sec:introduction}
Interference is a key bottleneck in wireless communication networks of all types: whenever spectrum is shared between multiple users, each user must deal with undesired signals. Cellular networks in densely populated areas, for example, are severely limited by interference. To address this problem, the research community as well as the wireless communications industry have invested a great deal of effort in trying to develop efficient communication schemes to deal with interference.
Nevertheless, the current state-of-the-art systems rely on two basic approaches: either orthogonalizing the communication links across time or frequency, or sharing the same resource while treating other users' signals as noise. If there are $K$ co-located users, these approaches result in a fraction $1/K$ of the total resource being available to each user: performance severely degrades as the number of users increases. 
Interference alignment is one recent development (among others, such as hierarchical MIMO \cite{OLT07}) that has opened the possibility of significantly better
performance in interference-limited communications than traditionally thought
possible.

The basic idea of interference alignment is to align, or overlap, multiple interfering signals at each
receiver in order to reduce the effective interference.
Interference alignment has been used in the index coding literature since the late 90's, seemingly for the first time by Birk and Kol \cite{BK98}. For wireless communication, interference alignment was used by Maddah-Ali et al.
\cite{MMK08} and made more explicit by Jafar and Shamai \cite{JS08}, both for the multiple-input multiple-output (MIMO) X channel. 
But the extent of the potential benefit of interference alignment was first observed by Cadambe and Jafar \cite{CJ08} in application to the $K$-user interference channel, when they showed that for time-varying or frequency selective channels with unbounded diversity, $\frac K2$ total degrees of freedom are achievable using a basic linear precoding scheme.  In other words, somewhat amazingly, each user gets the same degrees of freedom as with only \emph{two} users in the system, independent of the total number of users~$K$. 
 The number of degrees of freedom in a system, defined later, is given by the total capacity normalized by the capacity of a single point-to-point link, in the limit of high signal-to-noise ratios (SNR). 

The $\frac K2$ result of Cadambe and Jafar requires \emph{unbounded} channel diversity, and it is unclear what the implication is for real systems with \emph{finite} channel diversity. Despite major effort by researchers over the last six or so years, little is known about how diversity affects the ability to align interference. Partial progress in this direction includes~\cite{BT09} for the three-user channel, as well as~\cite{sun2013two} for single-beam strategies.

Many practical systems are also equipped with multiple antennas and thus have \emph{spatial diversity}.
Multiple antennas are known to greatly increase the degrees of freedom of point-to-point systems.
In this paper we focus on how spatial diversity helps to deal with interference
by studying interference alignment for the MIMO interference channel. In order to focus on the effect of spatial diversity, we assume there is no time or frequency diversity, i.e., the channel is constant over time and frequency. For technical reasons we additionally restrict attention to strategies making use of a single time-slot (no symbol extensions).

Because \cite{CJ08} uses linear (vector space) precoding, we attempt to simplify matters by restricting to the class of such vector space schemes (defined carefully in Sec.~\ref{sec:results}).
Our first results are for the symmetric three user channel. We prove a necessary condition for alignment and give a constructive achievable vector space strategy. These together characterize the feasibility of alignment for $d$ signaling dimensions per user in a symmetric system with $M$ antennas at the transmitters and $N$ antennas at the receivers. The arguments require only basic linear algebra.

Next we generalize to an arbitrary number of users. The arguments make use of tools from algebraic geometry to analyze the bilinear equations arising from the alignment conditions.
We first prove a general necessary condition. We subsequently show using
Schubert calculus that if $M=N$ and all transmitters use $d$ signaling dimensions, then the necessary condition is also sufficient. Schubert calculus is a system for computing the number of solutions to various enumerative problems on algebraic varieties such as Grassmannians. The framework gives an explicit (albeit complicated) combinatorial rule for counting the number of alignment solutions, and in principle allows to check directly if alignment is generically feasible. How to perform this verification is not obvious in general; we argue that the number of solutions is positive in the aforementioned symmetric case.

The rest of the paper is outlined as follows. In Section~\ref{sec:results} we introduce the MIMO interference channel model and formulate the alignment feasibility problem. Section~\ref{sec:mainResults} contains an overview of the main results and comparison with related works. 
Section~\ref{sec:necessary} derives necessary conditions for alignment, and Section~\ref{sec:sufficient} gives sufficient conditions. Finally, Section~\ref{sec:computing} discusses computing the number of alignment solutions as well as how to compute the solutions themselves,
and Section~\ref{sec:conclusion} gives some concluding remarks.

\section{The MIMO interference channel}
\label{sec:results}

The $K$-user MIMO interference channel has $K$ transmitters and $K$ receivers, with transmitter $i$ having $M_i$ antennas and receiver $i$ having $N_i$ antennas. For $i=1,\dots,K$, receiver $i$ wishes to obtain a message from the corresponding transmitter $i$. The remaining signals from transmitters $j\neq i$ are undesired interference.
The channel is assumed to be constant over time, and at each time-step the input-output relationship is given by
\begin{equation}
  \y_i=\bH{i}{i}\x_i+\doublesum{1\leq j\leq K}{j\neq i}\bH{i}{j}\x_j+\z_i\,,\quad 1\leq i\leq K\,.
\end{equation}
Here for each user $i$ we have $\x_i\in \C^{M_i}$ and $\y_i, \z_i\in \C^{N_i}$, with $\x_i$ the transmitted signal, $\y_i$ the received signal,  and $\z_i\sim \CN(0,I_{N_i})$ is additive isotropic white Gaussian noise. The channel matrices are given by $\bH{i}{j}\in \C^{N_i\times M_j}$ for $1\leq i,j\leq K$; for the rest of the paper, we assume
that the $\bH{i}{j}$ are generic, meaning that their entries lie outside of
an algebraic hypersurface depending only on the parameters $d_i$, $M_i$,
and~$N_i$.
If the entries are randomly chosen from some
non-singular probability distribution, this will be true with probability~$1$.
Additionally, each user obeys an average power constraint, $\frac1T \E(||\x_i^T||^2)\leq P$ for a block of length~$T$.

We restrict the class of coding strategies to (linear) \emph{vector space} strategies.
In this context, degrees-of-freedom has a simple interpretation as the
dimensions of the transmit subspaces, described in the next paragraph. However,
note that one can more generally define the degrees-of-freedom region in terms
of an appropriate high transmit-power limit $P\to \infty$ of the Shannon
capacity region $C(P)$ normalized by $\log P$ (\cite{MMK08,CJ08}). In that
general framework, it is well-known and straightforward that vector space strategies give a concrete non-optimal achievable strategy with
rates $$R_i(P)=d_i\log(P)+O(1), \quad 1\leq i\leq K\,.$$ Here $d_i$ is the dimension of transmitter $i$'s subspace and $P$ is the transmit power.

The transmitters encode their data using vector space precoding.
Suppose transmitter $j$ wishes to transmit a vector $\hat x_j \in \mathbb
C^{d_j}$ of
$d_j$ data symbols. These data symbols are modulated on the subspace
$U_j\subseteq \C^{M_j}$ of dimension~$d_j$, giving the input signal
$\bU_j\hat x_j$, where $ \bU_j$ is a $M_j \times d_j$ matrix whose
columns give a basis of~$U_j$. This signal is observed by receiver $i$ through the channel
as $\bH{i}{j} \bU_j\hat x_j$.
The dimension of the transmit
space, $d_j$, determines the number of data streams, or degrees-of-freedom,
available to transmitter $j$. With this restriction to vector space strategies,
the output for receiver~$i$ is
given by
\begin{equation}
  \y_i=\sum_{1\leq j\leq K}\bH{i}{j} \bU_j\hat x_j+\z_i\,,\quad 1\leq i\leq K\,.
\end{equation}
The desired signal space at receiver $i$ is thus $\bH{i}{i}U_i$, while the
interference space is the span of the undesired subspaces, i.e.,\ $\sum_{j\neq
i}\bH{i}{j}U_j$.

In the regime of asymptotically high transmit powers, in order that decoding can
be accomplished we impose the constraint at each receiver~$i$ that the desired
signal space $\bH{i}{i}U_i$ is complementary to the interference space
$\sum_{j\neq i}\bH{i}{j}U_j$.  Equivalently, receiver $i$ must have a subspace~$V_i$ (onto which it can project the received signal) with $\dim V_i=
\dim U_i$ such that
\begin{equation}\label{e:Generalorthogonality}
 \bH{i}{j}U_j \perp V_i\,, \quad 1\leq i,j \leq K, \quad i\neq j\,,
\end{equation}
 and
\begin{equation}\label{e:projCond}
\dim(\text{Proj}_{V_i}\bH{i}{i}U_i)=\dim U_i\,.
\end{equation}
Here, $\bH{i}{j}U_j\perp V_i$ means that the two vector spaces are orthogonal
with respect to the standard Hermitian form on $\C^{N_i}$.
Equivalently, if we write $\bV_i$ and $\bU_j$ for matrices whose columns form
bases for $V_i$ and~$U_j$ respectively, then orthogonality
means that all entries of the matrix $ \bV_i^\dagger \bH{i}{j} \bU_j$ are zero, where
$\bV_i^\dagger$ denotes the Hermitian transpose.

If each direct channel matrix $\bH{i}{i}$
has generic (or i.i.d.\ continuously distributed) entries, then the second condition~\eqref{e:projCond} is satisfied assuming $\dim V_i=d_i$ for each~$i$. This is because the set of channels for which condition~\eqref{e:projCond} is \emph{not} satisfied obeys a determinant equation and is therefore contained in an algebraic hypersurface (algebraic set of codimension 1).   This can be easily justified---see \cite{GCJ08} for some brief remarks. Hence we focus on condition \eqref{e:Generalorthogonality}.

Our goal in this paper is to determine when interference alignment is feasible:
given a number of users~$K$, numbers of antennas $M_1, \ldots, M_K$ and $N_1,
\ldots, N_K$, and desired transmit subspace dimensions $d_1, \ldots, d_K$, does
there exist a choice of subspaces $U_1, \ldots, U_K$ and $V_1, \ldots, V_K$ with
$\dim U_i = \dim V_i = d_i$ satisfying (\ref{e:Generalorthogonality})?

\section{Main results}
\label{sec:mainResults}
As discussed in the previous section, for vector space strategies the alignment problem reduces to finding
vector spaces $U_i \subset \C^{M_i}$ and $V_i \subset \C^{N_i}$ where $\dim U_i
= \dim V_i$ is denoted $d_i$, such that
\begin{equation}\label{e:GeneralorthogonalityIntro}
\bH{i}{j}U_j \perp V_i\,, \quad 1\leq i,j \leq K, \quad i\neq j\,,
\end{equation}
where the matrix $\bH{i}{j} \in \C^{N_i \times M_{j}}$ represents the channel
between transmitter~$j$ and receiver~$i$. We again emphasize that the $\bH{i}{j}$ are assumed to be generic. 

Our goal is to maximize the signal dimensions $d_i$ subject to the constraint
that there exist vector spaces satisfying~(\ref{e:GeneralorthogonalityIntro}).
Yetis et al. \cite{YGJK10} proposed comparing the number of variables and equations in the system of bilinear equations \eqref{e:GeneralorthogonalityIntro} in order to determine when it has solutions, and justified this using Bernstein's Theorem in the case that $d_i=1$ for all $i$. 
Our first result makes this intuition precise by showing that the feasible solutions are an algebraic
variety of the ``expected'' dimension, when the channel matrices are generic.
Thus, we have the following necessary
condition for interference alignment:

\begin{theorem}\label{t:upper-bound}
Fix an integer $K$ and integers $d_i$, $M_i$, and~$N_i$ for $1 \leq i \leq K$
and suppose the channel matrices $\bH{i}{j}$ are generic. If, for any subset $A
\subset \{1, \ldots, K\}$, the quantity
\begin{equation*}
t_A = \sum_{i\in A} \big(d_i(N_i - d_i) +  d_i(M_i - d_i)\big)
- \sum_{i, j \in A, i \neq j} d_i d_j.
\end{equation*}
is negative, then there are no feasible strategies. Moreover, if there are
feasible strategies, then $t_{\{1, \ldots, K\}}$ is the dimension of the
variety of solutions.
\end{theorem}

The constraint on $t_{\{1, \ldots, K\}}$ was obtained independently and
simultaneously by Razaviyayn et al.~\cite{RGL11}.

The dimension of the variety of solutions is important, because when multiple
strategies are feasible, we may wish to optimize over the feasible strategies
according to some other criterion, such as the robustness of the system.

The necessary condition from Theorem~\ref{t:upper-bound} is not sufficient. One,
almost trivial, requirement for there to even exist vector spaces is that $d_i
\leq
M_i$ and $d_i \leq N_i$ for each~$i$. Similarly, by looking at the total
capacity of two transmitters and one receiver, we have the following constraint,
which is also implied by combining the information-theoretic arguments
of~\cite{GJ10} and~\cite{JF07}.

\begin{theorem}\label{t:3-user-necessary-r-1}
For any distinct indices $i$, $j$, and~$k$, the feasibility of interference
alignment requires
\begin{equation*}
d_i + d_j + d_k \leq \max(N_i, M_j + M_k)
\end{equation*}
\end{theorem}

Symmetrically, Theorem~\ref{t:3-user-necessary-r-1} also holds with the roles
of~$M$ and~$N$ reversed. Moreover, this result extends to chains of transmitters
and receivers longer than three, at least when the transmit and receive
dimensions are identical. To do so, rather than three indices $i$, $j$, and~$k$
from Theorem~\ref{t:3-user-necessary-r-1}, we use a
sequence of indices such that each consecutive triple of indices is distinct.
Again, Theorem~\ref{t:3-user-necessary} also applies with $M$ and~$N$ reversed.

\begin{theorem}\label{t:3-user-necessary}
Fix a non-negative integer $r$ and let $i_1, \ldots, i_{r+2}$ be a sequence of
indices, such that each consecutive triple consists of three distinct indices,
i.e.,\ $i_j \neq i_{j+1}$
for $1 \leq j \leq r+1$ and $i_j \neq i_{j+2}$ for $1 \leq j \leq r$.
Also, assume that if $i_j = i_{j'}$ then $i_{j+1} \neq i_{j'+2}$.

Suppose that $N_{i_j}$ is the
same for all $1 \leq j \leq r$, which we denote $N$, and similarly $M_{i_j} = M$
for $2 \leq j \leq r+2$.
In order for interference alignment to be feasible, we must have:
\begin{equation*}
\sum_{j=1}^r d_{i_j} + \sum_{j=2}^{r+2} d_{i_j}
\leq \max(rN, (r+1)M)
\end{equation*}
When $r$ is positive, we can rewrite the left-hand side:
\begin{equation*}
d_{i_1} + d_{i_{r+1}} + d_{i_{r+2}}
+ 2\sum_{j=2}^r d_{i_r}
\leq \max(rN, (r+1)M)
\end{equation*}
\end{theorem}

\begin{figure}[tb]
\begin{center}
\includegraphics[width=\figwidth]{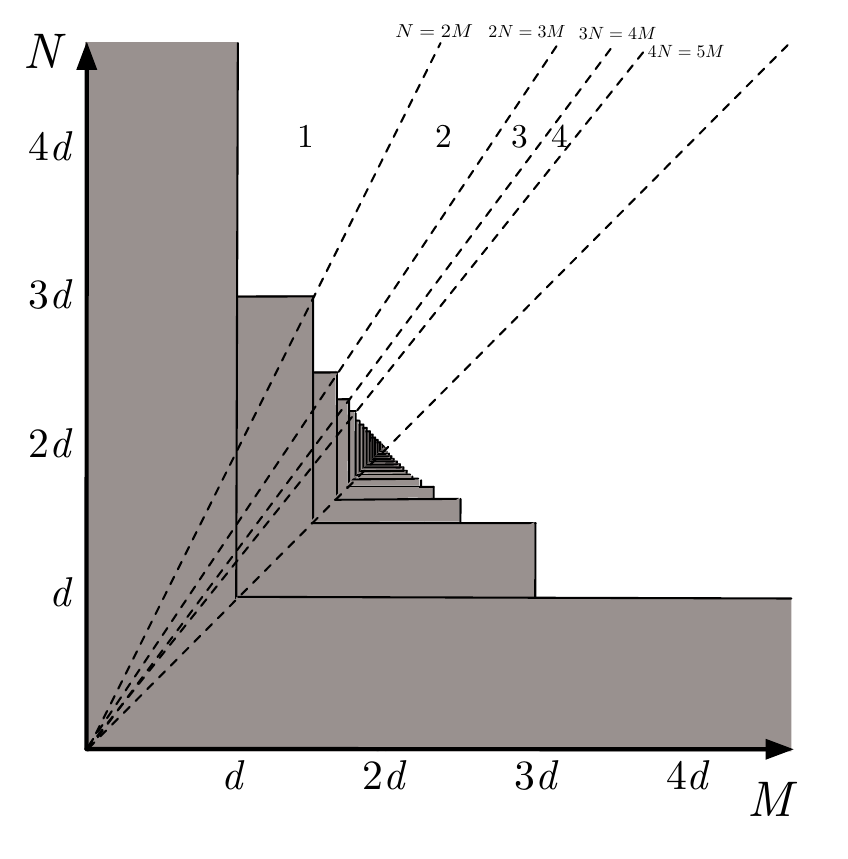} 
\vspace{-0.9cm}
\caption{For a fixed value of $d$, the feasible region in the $M,N$ plane is white while the infeasible region is shaded. 
The labels $1,2,3,4,\ldots$ indicate the maximum length of alignment paths for $M,N$ in the corresponding region.}
\label{f:3userfeasibleRegionINTRO}
\end{center}
\vspace{-.6cm}
\end{figure}

The condition on the indices~$i_j$ in Theorem~\ref{t:3-user-necessary} is
somewhat technical, but the simplest choice is to have $i_1, \ldots,
i_{r+2}$ cycle through three distinct values. Indeed, in the case of 3 users such a cycle is the only
possible sequence and we obtain the following simplification.

\begin{corollary}\label{c:necessary-specialized-3-user}
Suppose that $K=3$ and that $N_1 = N_2 = N_3 = N$ and $M_1 = M_2 = M_3 = M$. If
interference alignment is feasible, then for any positive integer~$r$,
\begin{equation*}
d_1 + d_{r+1} + d_{r+2} + 2 \sum_{j=2}^r d_r \leq \max(rN, (r+1) M),
\end{equation*}
where for $i > 3$, we define $d_i$ to be $d_{\overline i}$, where $\overline i$
is the remainder of~$i$ when divided by~$3$.
\end{corollary}
Of course, Corollary~\ref{c:necessary-specialized-3-user} also holds for any
reordering of the three users.
Moreover,
if we assume that the transmit dimensions~$d_i$ are all equal to some fixed~$d$, then
the infeasible parameters correspond to the shaded areas of
Figure~\ref{f:3userfeasibleRegionINTRO}. In fact, as shown by our next theorem, in this symmetric case we
have completely characterized the feasible region.

\begin{theorem}\label{t:3-user-sufficient}
Let $K=3$ and without loss of generality, let us assume that $N \geq M$. Then interference alignment is
feasible if and only if for each $r \geq 0$,
\begin{equation}\label{e:feasibility}
(2r + 1) d \leq \max(rN, (r+1)M)\,.
\end{equation}
Moreover, if, in addition, $N + M = 4d$, then there is a unique solution if $N >
M$ and there are $\binom{2d}{d}$ solutions if $N = M = 2d$.
\end{theorem}

The sufficiency of these conditions has also been independently
shown by Amir et al.~\cite{AEN11}, but only in the zero-dimensional case, when
$N+M = 4d$.
Also independently and simultaneously, Wang et al.~\cite{WGJ11} have obtained results very similar to Theorem~\ref{t:3-user-sufficient}. Their necessary condition (which matches the one in Theorem~\ref{t:3-user-sufficient}) is
information-theoretic, and thus, unlike ours, is not limited to linear strategies, constant channels, or no symbol-extensions. The linear achievability result of~\cite{WGJ11} also matches Theorem~\ref{t:3-user-sufficient}.

Theorem~\ref{t:3-user-sufficient} is constructive in the sense that the solutions
can be obtained from basic operations in linear algebra. The reason that the
$K=3$ case is easy to analyze is that the
constraints~(\ref{e:GeneralorthogonalityIntro}), which always link a pair of
vector space choices, form a cycle in the case of $K=3$. This cycle allows
alignment strategies to be constructed from alignment paths, as was also
observed by~\cite{WGJ11} (and previously in~\cite{BT09} for the three-user
interference channel with either time or frequency diversity).
When there are more users, we nonetheless have a sufficiency condition in the
case when $N=M$:

\begin{theorem}\label{t:symmetric}
Suppose that $K \geq 3$ and that $d_i = d$ and $M_i = N_i = N$ for
all users $i$. Then, for generic channel matrices, there is a feasible strategy
if and only if $2N \geq (K+1)d$.
\end{theorem}

A very similar theorem was also obtained by Razaviyayn et al.~\cite{RGL11} for the case
that $d$ divides $N$. They found a distinct, but overlapping, set of
parameters for which they could characterize feasibility.

\begin{theorem}[Theorem~2 in \cite{RGL11}]\label{t:divisible}
Suppose that $d = d_i$ for all $i$ and that $M_i$ and $N_i$ are divisible by $d$
for all~$i$. Then interference alignment is possible if and only if the
quantities $t_A$ from Theorem~\ref{t:upper-bound} are non-negative for all
subsets~$A$.
\end{theorem}

Rearranging the inequality of Theorem~\ref{t:symmetric}, we have that the
number of transmit dimensions satisfies $d\leq \frac{2N}{K+1}$.

\begin{corollary}[Fully symmetric achievable dof]\label{c:fully-symmetric-dof}
The maximum normalized dof is given by
\begin{equation*}
\text{max dof} = \frac{K}{N}\left\lf\frac{2N}{K+1}  \right\rf\leq 2\frac K{K+1}
\leq 2.
\end{equation*}
\end{corollary}

In sharp contrast to the $\frac K2$ total normalized dof achievable for infinitely many parallel channels in~\cite{CJ08}, for the MIMO case we see that at most \emph{two} dof (normalized by the single-user performance of $N$ transmit dimensions) are achievable for any number of users $K$ and antennas~$N$. The difference is due to the structure of the channel matrices: for the MIMO case with full generic matrices condition~\eqref{e:Generalorthogonality} is difficult to satisfy and \eqref{e:projCond} is easy, while for the parallel case with diagonal matrices the situation is reversed. 
We note that this observation was previously made in \cite{YGJK10} based on their conjectured necessary condition. 

Theorem~\ref{t:symmetric} suggests an engineering interpretation for
the performance gain from increasing the number of antennas. Depending on
whether $N< d(K+1)/2$ or not, there are two different regimes for the
performance benefits of
increasing $N$: (1) \emph{alignment gain} or (2) \emph{MIMO gain}. To illustrate these concepts, suppose that
there are $K=5$ users.
If $N=1$, i.e., there is only a single antenna at each
node, then no alignment can be done and only one user can communicate on a
single dimension, giving 1 total degree of freedom.
As the number of antennas increases to $2$ and~$3$, the number of degrees of
freedom becomes $3$ and~$5$ respectively. Because of alignment,
the \emph{number of users communicating} and hence total degrees of freedom increases, which we
call \emph{alignment gain}. The alignment gain has slope~$2$. However, after $N=3$, increasing $N$ affords no
additional possibilities for interference alignment; the total number of
degrees of freedom increases only because more dimensions are available. The \emph{MIMO
gain} has slope $5/3$, which is the asymptotic coefficient in
Corollary~\ref{c:fully-symmetric-dof}.

Unlike Theorem~\ref{t:3-user-sufficient}, the proof of Theorem~\ref{t:symmetric} does
not provide a way of computing the solutions. Instead of linear algebra, it
uses Schubert calculus to prove the existence of solutions. In fact, in Section~\ref{sec:computing}, we will see
that there cannot be a simple, exact description of the symmetric interference
alignment problem. Nonetheless, as we will
discuss, solutions may be found using numerical algebraic geometry software.

In addition to general algebraic methods of root finding, others have proposed heuristic
algorithms, mainly iterative in
nature (see~\cite{GCJ11}, \cite{PH09}, \cite{RSL10}, \cite{PD10},
and~\cite{SGHP10}). Some have proofs of convergence, but no performance
guarantees are known.
Schmidt et al. \cite{SGHP10}, \cite{SUH10} study a refined version of the
single-transmit dimension problem, where for the case that alignment is
possible,
they attempt to choose a good solution among the many possible solutions. Papailiopoulos and Dimakis \cite{PD10} relax the problem of maximizing degrees of freedom to that of a constrained rank minimization and propose an iterative algorithm.

In terms of the feasibility problem, Gonz\'alez, Beltran, and
Santamar\'ia~\cite{GBS14}
have given a polynomial-time randomized algorithm for determing whether given
parameters are feasible.
In a slightly different direction, Razaviyayn et al.~\cite{RSL10} show that
checking the feasibility of alignment for specific system parameters, including
the channel matrices, is NP-hard.
Note that their result does not contradict that of Gonz\'alez et al., since the
NP-hardness reduction requires special choices for the channel matrices and does not apply to generic channels.

Finally, we emphasize that our attention has been restricted to vector
space interference alignment, where the effect of finite channel diversity can
be more easily observed. Interfering signals can also be aligned on the signal scale using
lattice codes, which was first proposed in~\cite{BPT10} with followup work
in~\cite{CJS09}, \cite{EO09}, and~\cite{MGMK09}. Recent progress in this direction includes~\cite{OE11}, \cite{WSV11}, \cite{MN12}, and~\cite{OEN13}.

\section{Necessary conditions}
\label{sec:necessary}
In this section, we prove the necessary conditions for interference
alignment, Theorems~\ref{t:upper-bound}, \ref{t:3-user-necessary},
and~\ref{t:3-user-necessary-r-1}. The proof
of Theorem~\ref{t:upper-bound}
is by ``counting equations,'' or, more precisely, by determining
the dimension of the relevant algebraic varieties. The proofs of
Theorems \ref{t:3-user-necessary} and~\ref{t:3-user-necessary-r-1} use only
linear algebra. For background on some of the concepts in algebraic geometry, see
the texts by Hartshorne~\cite{Hartshorne} or Shafarevich~\cite{Shaf}.

In practice, interference alignment requires finding the feasible communications
strategies given the fixed channel matrices.
However, it will be useful to think of the procedure in reverse: we fix
the communication strategy and study the set of channels for which the
communication strategy is feasible. As we will see in the proof of
Lemma~\ref{l:dim-x}, the advantage here is that
for a fixed strategy, the constraints on the channel matrices are linear.

To make this approach precise, we will represent the space of strategies as a
product of Grassmannians.
Recall (for example from~\cite{Shaf}) that the \emph{Grassmannian} $G(d,N)$ is
the variety whose points correspond to $d$-dimensional subspaces of an
$N$-dimensional vector space~$\C^N$.
Thus, for each~$i$, the transmit subspace~$U_i$ corresponds to a point
in the Grassmannian, which we also write as $U_i\in G(d_i,M_i)$, and similarly
$\overline V_i\in G(d_i,N_i)$, where $\overline V_i$ is the complex conjugate
of~$V_i$. We choose to parametrize by the complex conjugate because the
relation~(\ref{e:GeneralorthogonalityIntro}) is defined by algebraic equations
in the basis of $\overline V_i$,
but not in $V_i$.
The strategy space is thus the product of the Grassmannians,
\begin{equation*}
\calS=\prod_{i=1}^KG(d_i,M_i)\times  \prod_{i=1}^K G(d_i, N_i)\,.
\end{equation*}
Likewise, the relevant channel matrices are a tuple of $K(K-1)$ matrices, which we can
represent as a point in the product $\HH=\prod_{i\neq j} \C^{N_i \times M_j}$.

In the product $\mathcal S \times \mathcal H$, we define the \emph{alignment
variety}
to be the subvariety $\mathcal I\subseteq \calS\times \HH$
of those ordered pairs $(s, h)$ such that $s$ is a feasible strategy for $h$.
The dimensions of $\calS$ and $\HH$ are the sums of the dimensions of their
factors, so
\begin{equation}\label{e:dimS}
\dim \mathcal S = \sum_{i=1}^K \big(d_i (M_i - d_i) + d_i(N_i - d_i)\big)
\,,
\end{equation}
and
\begin{equation}\label{e:dimH}
\dim \mathcal H = \doublesum{1 \leq i,j \leq K}{i \neq j} M_i N_j\,.
\end{equation}
The dimension of $\mathcal I$ will be computed in Lemma~\ref{l:dim-x}, using
Theorem~\ref{t:fibers}, which is a
rough analogue of the rank-nullity theorem from linear algebra.

Given a map $f\colon X \to Y$, the \emph{fiber} of a  point $y \in Y$ is the
inverse image of $y$ under the map~$f$:
\begin{equation*} f\inv(y)=\{x \in X: f(x) = y\}.\end{equation*}
A \emph{polynomial map} is a function whose coordinates are given polynomials.
Finally, an \emph{irreducible} variety is one which cannot be written as the
union of two proper, closed subvarieties.
The following theorem in algebraic geometry can be found, for example, as
Theorem~7 on page~76 of~\cite{Shaf}.

\begin{theorem}[Dimension of fibers]\label{t:fibers}
Let $f\colon X\to Y$ be a polynomial map between irreducible varieties. Suppose
that $f$ is dominant, i.e., its image is dense in~$Y$. Let $n$ and~$m$ denote
the dimensions of $X$ and~$Y$ respectively. Then $m\leq n$ and:
\begin{enumerate}
\item For any $y\in f(X) \subset Y$ and for any component~$Z$ of the fiber
$f\inv (y)$ the dimension of $Z$ is at least  $n-m$.
\item There exists a nonempty open subset $U\subset Y$ such that $\dim f\inv (y)=n-m$ for $y\in U$.
\end{enumerate}
\end{theorem}
In the proof of Theorem~\ref{t:upper-bound}, we will apply
Theorem~\ref{t:fibers} twice, for each of the projections from~$\II$ to the
factors~$\mathcal S$ and~$\mathcal H$. The first of these projections computes
the dimension of~$\mathcal I$.

\begin{lemma}\label{l:dim-x}
$\mathcal I$ is an irreducible variety of dimension
\begin{equation*}
\sum_{i=1}^K \big(d_i (M_i - d_i) + d_i(N_i - d_i)\big)
+ \doublesum{1 \leq i,j \leq K}{i \neq j} (M_i N_j - d_i d_j)\,.
\end{equation*}
\end{lemma}

\begin{IEEEproof}
We consider the projection from our incidence variety on the space of strategies
$p\colon\mathcal I \rightarrow \mathcal S$.
For any point $s= (U_1, \ldots,
U_K, V_1, \ldots, V_K)\in \calS$, we claim that the fiber $p\inv (s)$ is  a
linear space of dimension
\begin{equation*}
\dim p\inv (s)=\doublesum{1 \leq i, j\leq K}{i \neq j}( M_i N_j - d_i d_j)\,.
\end{equation*}

To see this claim, we give local coordinates to each of the subspaces comprising the solution $s\in \calS$. We write $u^{(i)}_a$ for the $a$th basis element of subspace $U_i$, where $u^{(i)}_a$ has zeros in the first $d_i$ entries except for a 1 in the $a$th entry, and similarly for $v^{(j)}_b$ (this is without loss of generality). 
The orthogonality condition $V_j\perp \bH{j}{i}U_i$ can now be written as the condition $v^{(j)}_b\perp \bH{j}{i} u^{(i)}_a$ for each $1\leq a\leq d_i$ and $1\leq b\leq d_j$. Writing this out explicitly, we obtain
\begin{align*}
0&=
\doublesum{1\leq k\leq M_i}{1\leq l\leq N_j} \overline v_b^{(j)}(k)\bH{j}{i}(k,l)u_a^{(i)}(l)
\\
&= \doublesum{1\leq k\leq d_i}{1\leq l\leq d_j}
\overline v_b^{(j)}(k)\bH{j}{i}(k,l)u_a^{(i)}(l) \\
&\qquad+\sum_{k> d_i \text{ or }  l>d_j} \overline v_b^{(j)}(k)\bH{j}{i}(k,l)u_a^{(i)}(l)
\\&= \bH{i}{j}(a,b)+\sum_{k> d_i \text{ or }  l>d_j} \overline v_b^{(j)}(k)\bH{j}{i}(k,l)u_a^{(i)}(l)\,.\label{e:orthogonality}
\end{align*}
Note that this equation is linear in the entries of $\bH{j}{i}$. There are $d_i
d_j$ such linear equations, and each one has a unique variable $\bH{j}{i}(a,b)$, so
the equations are linearly independent and each equation reduces the dimension
by 1. The claim follows from the fact that in total there are $\sum_{i\neq
j}d_id_j$ equations and we began with $\dim \HH =\doublesum{1 \leq i,j \leq K}{i \neq j} M_i N_j$ dimensions \eqref{e:dimH}.

We have shown that the fibers of $\mathcal I \rightarrow \mathcal S$ are vector
spaces, and, in particular, irreducible varieties of constant dimension. Thus,
since $\calS$ is an irreducible variety, so
is~$\mathcal I$~\cite[Ex.~14.3]{eisenbud}.
Moreover, Theorem~\ref{t:fibers} gives the relation
\begin{equation*}
\dim\II=\dim\calS+\dim p\inv(s)\,.
\end{equation*}
Since the dimension of~$\mathcal S$ is exactly the first
summation in the lemma statement, this proves the lemma.
\end{IEEEproof}

\begin{IEEEproof}[Proof of Theorem~\ref{t:upper-bound}]
We now consider the projection onto the second factor~$q\colon\mathcal I \rightarrow \mathcal H$.
This map is dominant if and only if the alignment problem is generically
feasible.
In this case,
Theorem~\ref{t:fibers} tells us that the fiber $q\inv (h)$ for a generic $h\in\mathcal
H$ has dimension \begin{equation}\label{e:dimqproj}
\dim q\inv (h)=\dim \mathcal I - \dim \mathcal H\,.\end{equation}
Using formula~(\ref{e:dimH}) for the dimension of $\mathcal H$ and
Lemma~\ref{l:dim-x} for the dimension of $\mathcal I$, we get that
(\ref{e:dimqproj}) is equal to the quantity~$t_{\{1, \ldots, K\}}$ from the
theorem statement.
Therefore, by Theorem~\ref{t:fibers}, this quantity must be non-negative if
there are to be feasible solutions, in which case $t_{\{1, \ldots, K\}}$ is the
dimension of the set of solutions to the generic alignment problem.

Now we turn to the necessary conditions for other subsets $A \subset \{1,
\ldots, K\}$.
Any feasible strategy for the full set of $K$ transmitters and
receivers, will, in particular be feasible for any subset of trasmitter-reciever
pairs. Therefore, a
necessary condition for a general set of channel matrices to have a feasible
strategy is that the same is true for any subset of the pairs. Since the
number~$t_A$ is the dimension of the variety of solutions when restricted just
to the
transmitters and receivers indexed by $i \in A$, then $t_A$ must be non-negative in order to have a feasible strategy.
\end{IEEEproof}

We now focus on the second necessary condition,
Theorem~\ref{t:3-user-necessary}, and the closely related
Theorem~\ref{t:3-user-necessary-r-1}. As we mentioned before,
Theorem~\ref{t:3-user-necessary} is a generalization of the obvious constraint
that $d_i \leq M_i$ for each transmitter, and the
generalization is formed by considering $r+1$ transmitters and $r$ receivers at
the same time. We first handle the case of $r=1$.

\begin{IEEEproof}[Proof of Theorem~\ref{t:3-user-necessary-r-1}]
We define $\bA$ to be the $N_{i} \times (M_{j} + M_{k})$ block matrix:
\begin{equation*}
\begin{pmatrix} \bH{i}{j} & \bH{i}{k} \end{pmatrix}.
\end{equation*}
For generic channel matrices, $\bA$ will have full rank, i.e., rank equal to
$\min(N_i, M_j + M_k)$. We consider the vector space $\UU = U_{j} \times U_k$ of
dimension $d_j + d_k$ in $\bC^{M_j + M_k}$. The orthogonality
condition~(\ref{e:Generalorthogonality}) implies that $V_i \perp \bA \UU$. If
$N_i \geq M_j + M_k$, then $\bA$ will be injective, and so $\bA\UU$ has
dimension $d_j + d_k$. However, orthogonal vector spaces can have at most
complementary dimensions, so we have that $d_i + d_j + d_k \leq N_i$.
On the other hand, if $N_i \leq M_j + M_k$, then the Hermitian transpose
$\bA^\dagger$ is injective, and from the orthogonality relation $\bA^\dagger V_i
\perp \UU$, we get that $d_i + d_j + d_k \leq M_j + M_k$. Thus, we conclude that
$d_i + d_j + d_k \leq \max(N_i, M_j + M_k)$.
\end{IEEEproof}

A key step in the proof of Theorem~\ref{t:3-user-necessary-r-1} was that
the matrix~$\bA$ had full rank. For $r > 1$, we again show that, under
appropriate hypotheses, the analogous matrix has full rank.

\begin{lemma}\label{l:AfullRank}
Let $i_1, \ldots, i_{r+2}$ be a sequence as in the statement of
Theorem~\ref{t:3-user-necessary}, and we assume that $N_i = N$ and $M_i = M$ for
all $i$.
For any $r\geq 1$ define the $rN\times (r+1)M$ block matrix~$\bA_r$ to be
\begin{equation}\label{e:big-matrix}
\mat{
\bH{i_1}{i_2} & \bH{i_1}{i_3} & & &
\\
 & \bH{i_2}{i_3} & \bH{i_2}{i_4} & &
 \\
& & \ddots & \ddots &
\\
& & & \bH{i_r}{i_{r+1}} & \bH{i_r}{i_{r+2}}
}.
\end{equation}
For generic channel matrices $\bH
ij$, the matrix $\bA_r$ has full rank, $\min(rN, (r+1)M)$.
\end{lemma}
Lemma~\ref{l:AfullRank} is proved in the appendix. Using the lemma, we now prove Theorem~\ref{t:3-user-necessary}.

\begin{IEEEproof}[Proof of Theorem~\ref{t:3-user-necessary}] 
We fix the integer~$r$.
Define the product of transmit spaces  $\UU=U_{i_2}\times U_{i_3}\times \cdots \times
U_{i_{r+2}}\subset(\C^M)^{r+1}$,
and similarly let $\VV=V_{i_1}\times\cdots \times V_{i_r}\subset (\C^N)^r$. Note that
$\UU$ and~$\VV$ have dimensions
\begin{equation*}
\dim \UU = \sum_{j=2}^{r+2} d_{i_j} \quad
\mbox{and}\quad
\dim \VV = \sum_{j=1}^{r} d_{i_j}.
\end{equation*}

First, suppose that $rN \geq (r+1)M$. Then Lemma~\ref{l:AfullRank} implies that
the linear map $\bA_r\colon (\C^M)^{r+1}\to(\C^N)^r$ is injective. By the
orthogonality condition~\eqref{e:Generalorthogonality},
we have $\VV\perp \bA_r \UU$, and thus
\begin{equation*}
\dim (\VV) + \dim(\bA_r \UU) =
\sum_{j=1}^r d_{i_j} + \sum_{j=2}^{r+2} d_{i_j}
\end{equation*}
is at most~$rN$.

Alternatively, if $(r+1)M\geq rN$, the Hermitian transpose $\bA_r^\dagger$ is
an injective linear map $\bA_r^\dagger\colon (\C^N)^r\to(\C^M)^{r+1}$. Again, the
orthogonality conditions~\eqref{e:Generalorthogonality} imply that $\bA_r^\dagger\VV
\perp \UU$ so $\dim \VV + \dim \UU \leq (r+1)M$. This proves the theorem.
\end{IEEEproof}

\section{Sufficient conditions}
\label{sec:sufficient}
In this section, we give criteria for ensuring the achievability of interference
alignment. We have already seen the necessary direction of
Theorem~\ref{t:3-user-sufficient}, and here we prove the sufficient direction, first
when $M=N$, and second when $M < N$. The former case will also be covered by
Theorem~\ref{t:symmetric} dealing with $K>3$, but we give a specific $K=3$ proof because it is
constructive and additionally it allows to compute the number of solutions in the boundary case. 

\subsection{Three users}
\begin{proposition}\label{p:k-3-symmetric}
If $K=3$, and $M=N \geq 2d$, then alignment is feasible. Moreover, in the case
of equality, the number of solutions is exactly $\binom{2d}{d}$ for generic
channel matrices.
\end{proposition}

We note that the construction in the following proof has also appeared
in~\cite[Appendix IV]{CJ08}.

\begin{IEEEproof}
By first restricting our transmit and receive spaces to arbitrary subspaces, we
can assume that $M=N=2d$.
Since the channel matrices are square, generically, they are invertible, so we
can define the product
\begin{equation*}
\mathbf{B} = \bH{1,}2 (\bH{3,}2)^{-1} \bH{3,}1 (\bH{2,}1)^{-1} \bH{2,}3 (\bH{1,}3)^{-1}.
\end{equation*}
Again, generically, this matrix will have $2d$ linearly independent
eigenvectors, and we
choose $V_1$ to be the span of any $d$ of them. Then we set
\begin{align*}
U_3^\perp &= (\bH{1,}3)^{-1} V_1 \\
V_2 &= \bH{2,}3 U_3^\perp \\
U_1^\perp &= (\bH{2,}1)^{-1} V_2 \\
V_3 &= \bH{3,}1 U_1^\perp \\
U_2^\perp &= (\bH{3,}2)^{-1} V_3.
\end{align*}
These form a feasible strategy, and there are $\binom{2d}{d}$ possible
strategies.
\end{IEEEproof}

Before we proceed to the proof in the case when $M$ and~$N$ are distinct, we informally
describe the geometry underlying the construction of solutions.

A given vector $u_i$ in the signal space of transmitter $i$ is said to initiate an alignment path of length $r+1$ if there exists a sequence of vectors $u_{i+1},u_{i+2},\dots,u_{i+r}\in \C^M$, such that
\begin{gather*}
\bH {i-1,}{i} u_i = \bH{i-1,}{i+1}u_{i+1}, \\
\vdots \\
\bH {i+r-2,}{i+r-1} u_{i+r-1}
= \bH{i+r-2,}{i+r} u_{i+r}.
\end{gather*}
Here channel indices are interpreted modulo~3. For example, a vector $u_2$ at transmitter 2 initiating an alignment path of length~3 means that there exist vectors~$u_3$ and~$u_1$ such that $\bH12 u_2 = \bH13 u_3$ and $\bH23 u_3 = \bH21 u_1$.

The feasible region of Figure~\ref{f:3userfeasibleRegionINTRO} is divided up into sub-regions labeled with the maximum length of an alignment path; this number depends on $M$ and $N$ through the incidence geometry of the images of the channel matrices~$\im(\bH ij)$.
We begin by examining sub-region~1, and then look at how things generalize to the other sub-regions.

\begin{figure}[bt]
\begin{center}
\includegraphics[width=\figwidth]{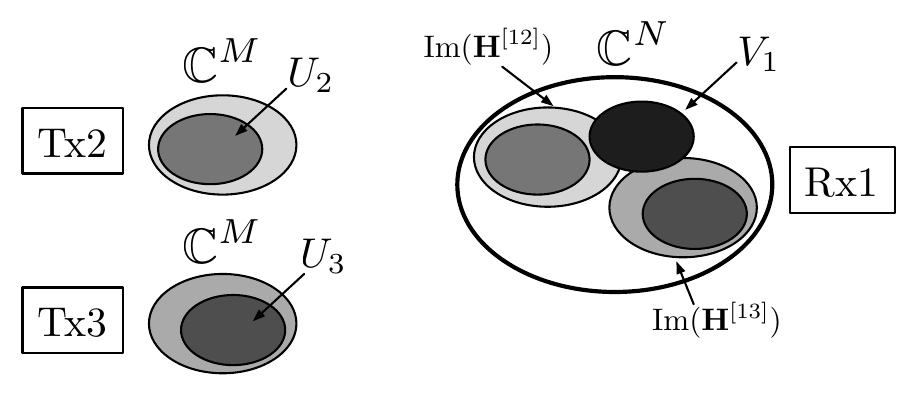}
\caption{Sub-region~1: The figure indicates that no alignment is possible when $2M\leq N$, since $\im(\bH12)$ and $\im(\bH13)$ are complementary. Since the three subspaces $V_1, \bH12 U_2,\bH13 U_3$ are each of dimension $d$, complementary, and lie in $\C^N$ at receiver 1, we obtain the constraint $3d\leq N$.  }
\label{f:NoAlignment}
\end{center}
\vspace{-.3cm}
\end{figure}

The point of departure is the obvious constraint $d\leq M$ in order to have a
$d$-dimensional subspace of an $M$ dimensional vector space. Continuing,
assuming $M\geq d$, suppose $ 2M\leq N$, so $(M,N)$ lies in sub-region~1 of
Figure~\ref{f:3userfeasibleRegionINTRO}. At receiver one, the images $\im(\bH12)$ and $\im(\bH13)$ of the channels from transmitters two and three are in general position and therefore their intersection has dimension $[2M-N]^+=0$; in other words, \emph{alignment is impossible} in sub-region~1. Figure~\ref{f:NoAlignment} shows pictorially that because alignment is not possible here, we have the constraint $3d\leq N$. 

\begin{figure}[tb]
\begin{center}
\includegraphics[width=\figwidth]{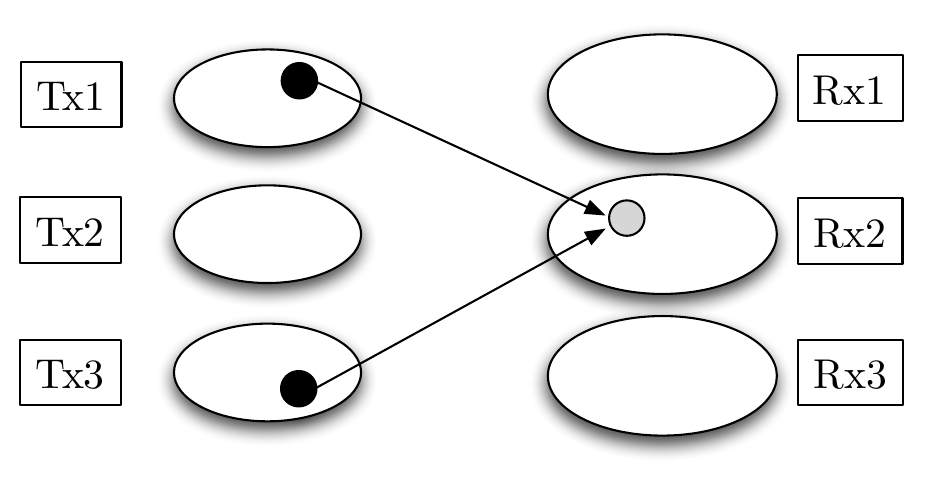}
\caption{Sub-region~2: Alignment is possible here. The figure denotes an alignment path of length~2.}
\label{f:AlignmentPath2}
\end{center}
\end{figure}

\begin{figure}[tb]
\begin{center}
\includegraphics[width=\figwidth]{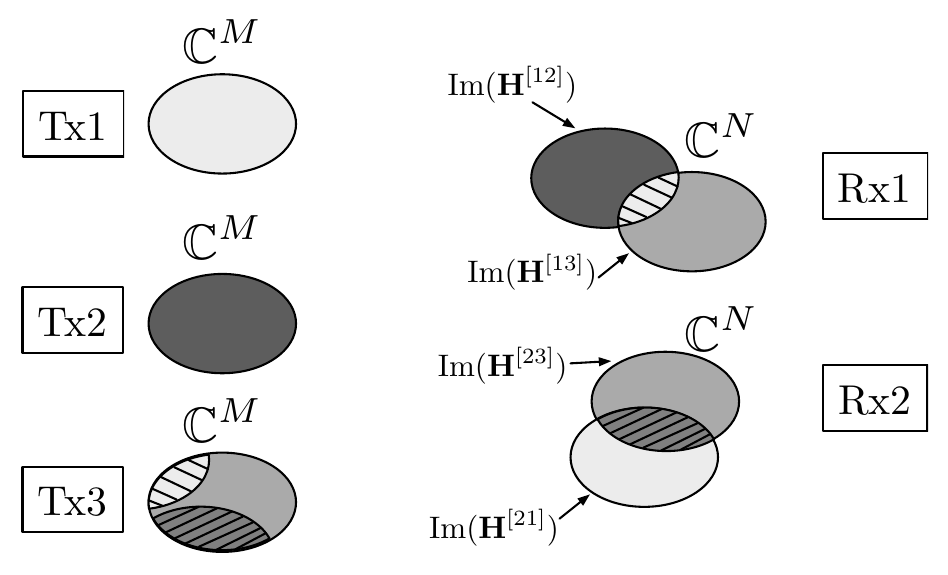}
\caption{Sub-region~2: The striped regions at receivers one and two each denote the dimension $2M-N$ portion of the space in which alignment can occur. From transmitter three's perspective, one sees that  \emph{simultaneous} alignment is not possible for $2(2M-N)\leq M$, or equivalently, $3M\leq 2N$. }
\label{f:NoSimulAlignment}
\end{center}
\vspace{-.3cm}
\end{figure}

Moving onward to sub-region~2, we have $2M> N$ and thus alignment \emph{is} possible. This means that alignment paths of length 2 are possible (Fig~\ref{f:AlignmentPath2}), with up to $2M-N$ interference dimensions overlapping at each receiver.
Thus, the interference space $\bH12 U_2 + \bH13 U_3$ at receiver one occupies at least $2d-(2M-N)$ dimensions, and we have the constraint $3d\leq 2M$.
However, because $3M\leq 2N$, no vector at (say) transmitter three can be \emph{simultaneously} aligned at both receivers one and two, as indicated in Figure~\ref{f:NoSimulAlignment}. One can also see that no simultaneous alignment is possible by changing perspective to that of a combined receiver one and two.
 By Lemma~\ref{l:AfullRank}, the map
\begin{equation}\label{e:Aexample}
\mat{\bH12 & \bH13 & \\ & \bH23 & \bH21}
\end{equation}
from the three transmitters to $\C^{2N}$ is injective; analogously to the case in sub-region 1, this is interpreted to mean that no alignment is possible in the combined receive space $\C^{2N}$.
Thus, five complementary $d$-dimensional  subspaces lie in $\C^{2N}$ and we obtain the constraint $5d\leq 2N$.

\begin{figure}[tb]
\begin{center}
\includegraphics[width=\figwidth]{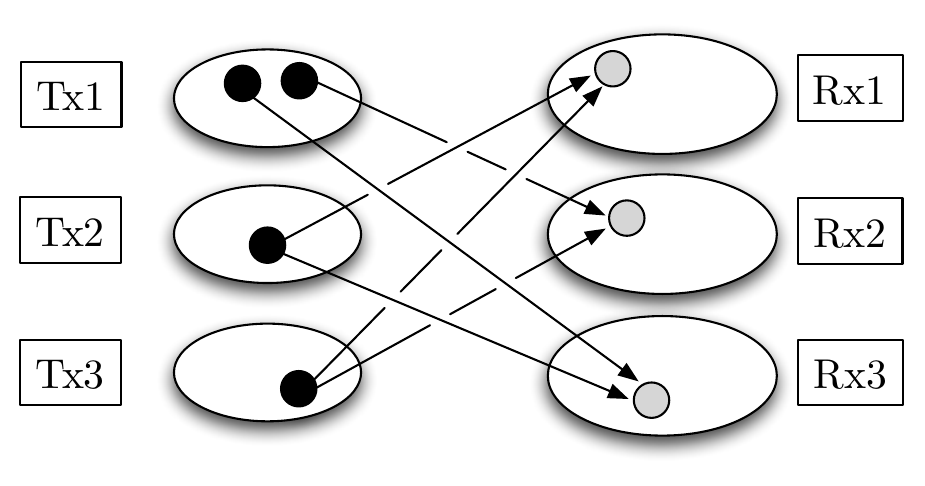}
\caption{Sub-region~4: Alignment paths of length four are denoted here, initiated by vectors at transmitter~1.}
\label{f:AlignmentPath4}
\end{center}
\vspace{-0.5cm}
\end{figure}

As far as achievability goes, the basic rule-of-thumb is to create alignment paths of maximum length. Thus, in sub-region~2, where alignment is possible, the achievable strategy aligns (as per Fig.~\ref{f:AlignmentPath2}) as many vectors as possible and the remaining ones (if $d>2(2M-N)$) are not aligned. For example, in sub-region~4, alignment paths of length four are used (Fig.~\ref{f:AlignmentPath4}).

\begin{proposition}\label{p:k-3-critical}
If $K=3$ and $M < N$ and (\ref{e:feasibility}) holds for each $r \geq 0$, then
alignment is feasible. In the $0$-dimensional case, when
$N+M=4d$, there is a unique solution.
\end{proposition}

\begin{IEEEproof}
Let $r$ be the (unique) integer such that
\begin{equation}
rN<(r+1)M\quad \text{and}\quad (r+1)N\geq (r+2)M\,.
\end{equation}
Note that this implies, from (\ref{e:feasibility}), that
\begin{equation}\label{e:N1}
(2r+3)d\leq (r+1)N
\end{equation}
and
\begin{equation}\label{e:M1}
(2r+1)d\leq (r+1)M\,.
\end{equation}

We prove achievability by examining two cases: first $d\leq (r+1)[(r+1)M-rN]$
and second, $d>(r+1)[(r+1)M-rN]$. Case 1 means that all of the signal space~$U_i$ can be obtained from alignment paths of length $r+1$ (up to integer rounding), whereas in case 2 we must use alignment paths of length $r$ as well in order to attain the required $d$ dimensions.

We first establish case 1. Let $\bA_r^i$ be the matrix
from~(\ref{e:big-matrix}) for the increasing sequence of indices $i,i+1,\dots,
i+r+1$, where indices are understood modulo~$3$.
Let $W_i$ be a dimension $\big\lf
\frac{d}{r+1}\big\rf$ subspace in the kernel of $\bA_r^i$. Let $d':=d-(r+1)\big\lf
\frac{d}{r+1}\big\rf$, and if $d'>0$ let $w_i$ be a 1-dimensional subspace in $\ker
\bA_r^i \setminus W_i$. We define $W_{i,j} \subset \C^M$ to be the projection
of $W_i$ onto its $(j-i)$th block of coordinates.
The spaces $w_i$ are required in order to accommodate the
remainder left when dividing $d$ by $r+1$, and will together contribute $d'$
dimensions to each signal space $U_j$.
We put
\begin{equation}\label{e:subspaceConstructionU}
U_j = \sum_{i=j-1}^{j-r-1}W_{i,j} + \sum_{i=j-1}^{j-d'}w_{i,j}
\end{equation}
and
\begin{equation}\label{e:subspaceConstructionV}
\begin{split}
V_j &= \Bigg( \bH{j,}{j-1}W_{j,j+1}+\bH{j,}{j-1}w_{j,j+1}\\
&\quad+\sum_{i=j}^{j-r}\bH{j,}{j+1} W_{i,j+1} + \sum_{i=j}^{j-d'+1}w_{i,j}\Bigg)^\perp,
\end{split}
\end{equation}
where again, the indices in $\bH{j,}{j-1}$ and $\bH{j,}{j+1}$ are understood
to be taken modulo three.

If all of $U_j$'s constituent subspaces are complementary, then $U_j$ has
dimension $(r+1)\big\lf\frac d{r+1}\big\rf+d'=d$. We rigorously justify this in
Lemma~\ref{l:UVdimension}.
To see that $V_j$ has dimension (at least) $d$, we observe that by subadditivity of dimension,
\begin{equation}
\dim V_j\geq N-(r+2)\left\lf \frac{d}{r+1}\right\rf-d' -e\,,
\end{equation} where $e=0$ if $(r+1)|d$ and $e=1$ otherwise.
Plugging in the inequality \eqref{e:N1} we obtain
\begin{align*}
\dim V_j &\geq \frac{2r+3}{r+1}d-d-e-\left\lf \frac{d}{r+1}\right\rf \\
&=d+\frac{d}{r+1}-\left\lf \frac{d}{r+1}\right\rf-e\geq d\,.
\end{align*}

Suppose now that we are in case 2: $d>(r+1)[(r+1)M-rN]$. This
means that not all of the signal space~$U_i$ can be included in alignment paths
of length $r+1$, so the remainder will be included in alignment paths of length $r$. Let $d':=d-(r+1)[(r+1)M-rN]$ and $d'' = d'-r\flr{\frac{d'}r}$.
As before, denote by $W_i$ the kernel of the matrix $\bA_r^i$, having dimension
$(r+1)M-rN$. Denote by $\pi$ the projection from $\C^{(r+1)M}\to\C^{rM}$ to the
first $rM$ coordinates. The space $\pi(\ker \bA_r^i)$ is contained in $\ker \bA_{r-1}^i$. Let $X_i$ for $i=1,2,3$ each be a $\flr{\frac{d'}{r}}$ dimensional subspace in $\ker \bA_{r-1}^i\setminus \pi(W_i)$, and let $w_i$ be a 1-dimensional subspace in $\ker \bA_{r-1}^i\setminus (\pi(W_i)+ X_i)$.
Put
\begin{equation}\label{e:subspaceConstruction2U}
U_j = \sum_{i=j-1}^{j-r-1}W_{i,j} + \sum_{i=j-1}^{j-r}X_{i,j} +\sum_{i=j-1}^{j-d''}w_{i,j}
\end{equation}
and
\begin{equation}\label{e:subspaceConstruction2V}\begin{split}
V_j = \Bigg( \bH{j,}{j-1}(W_{j,j+1}+X_{j,j+1}+w_{j,j+1}) \\
+\sum_{i=j}^{j-r}\bH{j,}{j+1} W_{i,j+1}+\sum_{i=j}^{j-r+1}\bH{j,}{j+1} X_{i,j+1} \\
+ \sum_{i=j}^{j-d'+1}w_{i,j}\Bigg)^\perp\!\!.
\end{split}\end{equation}
As before, a naive count suggests that $U_j$ should have dimension $d$, and this
will again be justified with Lemma~\ref{l:UVdimension}.

To see that $V_j$ has dimension at least $d$ we again use subadditivity of dimension to get
\begin{align*}
\dim V_j
&\geq N-(r+2)[(r+1)M-rN]-(r+1)\flr{\frac{d'}r} \\
&\qquad -d''-e_1 \\
&= N-(r+2)[(r+1)M-rN]- \flr{\frac{d'}r}\\
&\qquad -d'-e_1,
\end{align*}
where $e_1$ is $0$ if $r$ divides $d'$ and $e_1$ is $1$ otherwise. Letting $e_2:=\frac{d'}r - \flr{\frac{d'}r}$, we
have
\begin{align*}
\dim V_j&\geq N-(r+2)[(r+1)M-rN]-\frac{d'}r \\
&\qquad-d'+e_2-e_1 \\
&=N-\frac{(r+1)d}{r}+\frac{1}r[(r+1)M-rN] \\
&\qquad +e_2-e_1\\
&=d+\frac{(r+1)}{r}M-\frac{2r+1}{r}d+e_2-e_1
\,.
\end{align*}
Substituting $\frac{r+1}{2r+1}M$ for $d$, the inequality \eqref{e:M1} implies that
$$
\dim V_j\geq d+e_2-e_1\,.
$$
If $e_1$ is one then $e_2$ is strictly positive, so the fact that $\dim V_j$ is an integer implies $\dim V_j\geq d$.
\end{IEEEproof}

\begin{lemma}\label{l:UVdimension}
The subspaces $U_j$ and $V_j$ defined in (\ref{e:subspaceConstructionU}),
(\ref{e:subspaceConstructionV}), (\ref{e:subspaceConstruction2U}),
and~(\ref{e:subspaceConstruction2V}) have dimension $d$.
\end{lemma}
\begin{IEEEproof}
We first show that $U_1$ has dimension $d$; by symmetry of the construction, the dimensions of $U_2$ and $U_3$ will also be $d$.

The subspace $U_1=\sum_{i=-r}^0 W_{i,1}$ is the sum of $r+1$ subspaces $W_{i,j}$, which we claim are independent;
suppose to the contrary, that there is some set of linearly dependent vectors $w_{i_1},w_{i_2},\dots, w_{i_s}$, with $0\leq i_1\leq i_2\leq\dots\leq i_s\leq r$, and $w_i\in W_{-i,1}$, satisfying $w_{i_s}-\sum_{\ell=1}^{s-1} \lam_\ell w_{i_\ell} =0$. Let $s$ be the minimum such value, with all sets of subspaces $W_{i_1,j},W_{i_2,j},\dots,W_{i_{s-1},j}$ for $j=1,2,3$ being complementary.

Now, by the definition of the subspaces $W_{i,j}$, for each vector
$w_{i_\ell}\in W_{-i_\ell,1}$ there is a sequence $u^{2}_{i_\ell},\dots,u^{q+1}_{i_\ell}$ of length $q:=r+1-i_{s-1}$ satisfying $\bH{3}{1}w_{i_\ell}=\bH{3}{2}u^{2}_{i_\ell},\dots, \bH{q+2,}{q}u^{q}_{i_\ell}=\bH{q+2,}{q+1}u^{q+1}_{i_\ell}$. The linear combination $\sum_{\ell=1}^{s-1}\lam_\ell w_{i_\ell}$ thus gives rise to a sequence $u^1,\dots, u^{q+1}$ defined by $u^a = \sum_{\ell=1}^{s-1}\lam_\ell u^a_{i_\ell}$ satisfying
\begin{equation}\begin{split}\label{e:chain}
\bH{3}{1}w_{i_s}&=\bH{3}{1}\bigg(\sum_{\ell=1}^{s-1} \lam_\ell w_{i_\ell}\bigg) = \bH32 u^2,
\\
\bH12 u^2&=\bH13 u^3 \\
&\; \; \vdots
\\
\bH{q+2,}{q}u^{q}&=\bH{q+2,}{q+1}u^{q+1}\,.
\end{split}
\end{equation}
Note that by the minimality assumption of $s$, none of the $u^j$ vectors are zero.

By the definition of $W_{-i_s,1}$, there is a length-$(i_s-1)$ sequence of vectors preceding $w_{i_s}$ satisfying alignment conditions similar to those in \eqref{e:chain}; together with $w_{i_s}$ and the vectors in \eqref{e:chain}, this sequence can be extended to a sequence of vectors of total length $q+i_{s}=r+1+(i_s-i_{s-1})>r+1$, none of which are zero. Stacking the first $r+2$ of these vectors produces a nonzero element in the kernel of $\bA^{i_s}_{r+1}$. However, $\bA^{i_s}_{r+1}$ is full-rank by Lemma~\ref{l:AfullRank}; the dimension of the kernel is $\big[(r+2)M-(r+1)N\big]^+= M+d\big( 2r+1-2r-3\big)=M-2d<0$, i.e., the kernel is trivial. This is the desired contradiction.

We now check that $V_1$ has dimension $d$, and again by symmetry, the dimensions of $V_2$ and $V_3$ will also be $d$. Note that if $V_1$ had dimension greater than $d$, we could choose a $d$-dimensional subspace and this would still satisfy the alignment equations \eqref{e:Generalorthogonality}. But $V_1$ is the orthogonal complement of the sum of $r+2$ subspaces $W_{i,j}$ of dimension $d/(r+1)$, so by subadditivity of dimension, we have the lower bound on dimension $\dim V_1\geq N-(r+2)\dim W_{i,j}=d$.
\end{IEEEproof}

\subsection{More than three users}
We now prove achievability for more than 3 users. We do this under the
additional assumption that $M=N$. Unlike our techniques above, these
existence results will not be constructive, a characteristic shared by previous
existence proofs in~\cite{RGL11} and~\cite{BCT11a}.

In~\cite{BCT11a}, a proof of Theorem~\ref{t:symmetric} was given using
dimension theory for algebraic varieties and linear algebra. Here, we give an
intersection-theoretic proof, which involves more advanced machinery, but that
machinery allows for a more straightforward computation. Moreover, as we will see
in Proposition~\ref{p:numbers-solutions}, the Schubert calculus framework will also allow us
to go beyond the existence of solutions and count the number of solutions when
that number is finite.

Our proof will show that the expected number of solutions, as counted by
Schubert calculus, is always positive. Schubert calculus is a method for
computing the number of solutions to certain enumerative problems. For algebraic
varieties, such as the product of Grassmannians that parametrize alignment
strategies, there is a commutative ring whose elements correspond to conditions,
on the parameters (such as the orthogonality
relation~(\ref{e:GeneralorthogonalityIntro})), and where the product corresponds
to the simultaneous imposition of both conditions. In algebraic geometry, this
is known as the Chow ring and in the case of products of Grassmannians, it
coincides with the cohomology ring from algebraic topology.

The Chow ring of a Grassmannian has an explicit $\mathbb Z$-basis indexed by
partitions. Specifically, the Chow ring of the
Grassmannian $G(d, m)$ has a basis corresponding to partitions with at most $d$
parts of size at most $m-d$. Such a partition is a list of integers~$\lambda_i$
with $m-d \geq \lambda_1 \geq \cdots \geq \lambda_d \geq 0$, and we write
$\lvert \lambda \rvert$ for its size, $\lambda_1 + \cdots + \lambda_d$. The
product between two of these basis elements, known as Schubert classes, is given
by an intricate combinatorial process known as the Littlewood-Richardson rule.

Thus, to determine whether a given alignment problem is feasible, we proceed in
two steps. First, we determine the elements in the Chow ring corresponding to
each of the orthogonality conditions~(\ref{e:GeneralorthogonalityIntro}).
Second, we multiply these elements together and the resulting product is
non-zero if and only if alignment is feasible. The first step is done by
Lemma~\ref{l:containment-class} below. For the second step, the central
difficulty is understanding the results of the
Littlewood-Richardson rule for products of
Schubert classes. In order to establish sufficient conditions in the fully
symmetric case, our proof will use carefully chosen terms from each Schubert
class, which will be sufficient because the products of the other terms will be
non-negative:
\begin{theorem}\label{t:schubert-non-negativity}
In the Chow ring of the Grassmannian, the product of two partitions is a
non-negative sum of other
partitions~\cite[p.~146, (8)]{fulton}.
\end{theorem}

For the chosen terms, it will be sufficient to compute their
products, not with the general Littlewood-Richardson rule, but by the following
simplification:

\begin{proposition}\label{p:highest-weight-products}
The Schubert classes have the following properties in the Chow ring of the
Grassmannian:
\begin{enumerate}
\item If $\lambda$ and $\mu$ are two partitions with $\lambda_1 + \mu_1 \leq d$,
then the product $[\lambda][\mu]$ has a coefficient of $1$ in front of the
term~$[\nu]$, where $\nu_i = \lambda_i + \mu_i$.
\item Suppose that $\lambda$ has $\ell$ parts and $\mu$ has $k$ parts and that
$\ell + k \leq m-d$. Then $[\lambda][\mu]$ has a
coefficient of $1$ in front of the term~$[\nu]$ where $\nu$ is formed by
concatenating the parts of $\lambda$ with the parts of $\mu$, and then sorting
them in decreasing order.
\end{enumerate}
\end{proposition}

\begin{figure}
\begin{centering}
\includegraphics{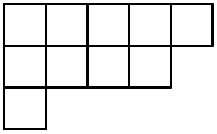}
\par\end{centering}
\caption{The Young diagram of the partition $(5, 4, 1)$.}
\label{f:partition}
\end{figure}
Both parts of this proposition can each be proved from the Pieri rule~\cite[p.
146, (9)]{fulton}, and in fact they are closely related to each other.
Partitions can be depicted as boxes in the upper left corner, such as
the depiction of $(5, 4,1)$ in Figure~\ref{f:partition}. Such diagrams have an
involution by reflecting them along a diagonal so that the conjugate of $(5,4,1)$
is the partition $(3,2,2,2,1)$. For any partition~$\lambda$, we write $\lambda'$
to denote the conjugate partition. This conjugation corresponds to the isomorphism
between $G(d, m)$ and $G(m-d, m)$, and it is compatible with the multiplication.
The last two items are related by this conjugation operation.

The Chow ring of a product of Grassmannians has a basis indexed by tuples of
partitions, one for each Grassmannian. Moreover, the products can be computed
factorwise.
Formally, the Chow ring of the product of Grassmannians is the tensor product of
the Chow rings of the Grassmannians.

We now compute the class in this Chow ring of the variety defined by a single
orthogonality condition~(\ref{e:GeneralorthogonalityIntro}). Our method is
similar to the elementary definition of the degree of a projective variety as
the number of points in the intersection with a generic linear space of
complementary dimension. However, instead of linear spaces, we look at the
intersection of Schubert classes of complementary dimension with the set of
pairs of vector spaces satisfying~(\ref{e:GeneralorthogonalityIntro}).

\begin{lemma}\label{l:containment-class}
The alignment correspondence defined by $\bH{i}{j}U_j \perp V_i$ has class
$\sum_\lambda [\lambda] \otimes [d^{d} - \lambda']$ in the Chow ring of
$G(d, M) \times G(d, N)$. The sum is taken to be over all
partitions with at most $d$ parts of size at most $d$, and $d^{d} -
\lambda'$ is the partition whose $k$th part has size $d - \lambda_{d + 1 -
k}'$.
\end{lemma}

\begin{IEEEproof}
We compute the class by intersecting with dual Schubert classes to get a
zero-dimensional cycle. In particular, we let $\mu$ and $\nu$ be partitions into
at most $d$ parts of size at most $M-d$ and $N-d$ respectively, and such that
the total size $\lvert \mu \rvert + \lvert \nu \rvert$ is $(M+N-3d)d$, which is
the dimension of the correspondence.

We first recall the definition of the Schubert variety in $G(d,M)$
associated to a partition $\mu$ and a flag $F_*$. A flag in $\C^M$ is a nested
set of vector spaces $0 = F_0 \subset F_1 \subset \cdots \subset F_M = \C^M$
such that $F_i$ has dimension~$i$. The Schubert variety is the closed subvariety
of those vector spaces~$U$ such that
\begin{equation*}
\dim(U \cap F_{M-d+i - \mu_i}) \geq i
\quad \mbox{for } 1 \leq i \leq d.
\end{equation*}

By symmetry, we can assume that $N$ is greater than or equal to~$M$ and thus a
vector space $U \in G(d, M)$ uniquely defines an $(N-d)$-dimensional subspace
$(\bH{i}{j} U)^\perp$ in~$\C^N$. Likewise, the orthogonal complement
$(\bH{i}{j}F_{k})^\perp \subset \C^N$ is an $(N-k)$-dimensional vector space,
which together form a flag from dimension $N-M$ through~$N$. We can choose a
flag of additional vector spaces contained within $(\bH{i}{j} \C^M)^\perp$ to
get a full flag in~$\C^N$, which we denote~$\widetilde F_{*}$. Then $U$ is in
the Schubert
variety of~$\mu$ if and only if $(\bH{i}{j}U)^\perp$ is in the
Schubert variety corresponding to $\widetilde F_*$ and the partition $((N-M)^d +
\mu)'$, which we denote~$\sigma$. Note that $\lvert \sigma \rvert = \lvert \mu
\rvert + (N-M)d$, so $\sigma$ and $\nu$ together have total size $(2M - 3d)d$.

We also fix a flag $E_*$ in~$\C^N$, which then defines a Schubert variety in
$G(d,N)$ indexed by~$\nu$. We assume that this flag~$E_*$ is chosen generically,
by which we mean that the intersection $\widetilde F_i \cap E_j$ is trivial if
$i + j \leq N$ and has the expected dimension $i + j - N$ otherwise.

Now we wish to find the points in the intersection of these two Schubert
varieties. Passing from $U$ to $\widetilde U = (\bH{i}{j} U)^\perp$ turns the
orthogonality condition into containment, so we wish to find pairs $V \subset
\widetilde U$ satisfying
the Schubert conditions
\begin{equation}\label{e:schubert-conditions}
\dim(\widetilde U \cap \widetilde F_{M-d+i - \nu_i}) \geq i,
\quad\dim( V \cap E_{d + j - \sigma_j}) \geq j,
\end{equation}
where $1 \leq i \leq d$ and $1 \leq j \leq N-d$.
We now let $i$ and~$j$ be indices satisfying $i + j = N - d + 1$.
Since $\widetilde U$ has dimension $N-d$ and contains $V$, this means that
$\widetilde U \cap \widetilde F_{M-d+i-\nu_i}$ and $V \cap E_{d+j-\sigma_j}$
must have a non-zero element in common, and thus
$\widetilde F_{M-d+i-\nu_i}$ and $E_{d+j+\sigma_j}$ must intersect
non-trivially. By the generic choice of the flag~$E_*$, this means
that the sum of the dimensions of these vector spaces must be at least the
dimension of the ambient space, so that we have the inequality $2M - d + 1 -
\nu_i - \sigma_j > M$.
Rearranging, this means that $\nu_i + \sigma_j$ can be
at most $M - d$. Because of their degrees, the only possibility is that $\nu_i =
M-d - \sigma_{M-d+1-i}$ for $1 \leq i \leq d$ and that $\sigma_j = d$ for $1
\leq j \leq M-2d$.

Moreover, for fixed partitions~$\nu$ and~$\sigma$ of this type, there is a unique
pair of vector spaces $V \subset \widetilde U$ satisfying
(\ref{e:schubert-conditions}). We set
$V$ to be the vector space generated by the one-dimensional vector spaces
$\widetilde F_{M-d + i - \nu_i} \cap E_{d - i + \mu_i+1}$ for $1 \leq i \leq d$
and take $\widetilde U$ to be generated by~$V$ and $\widetilde F_{M-2d}$.

What we have shown is that the only Schubert classes which occur in the
correspondence class are dual to the classes~$\mu$ and~$\sigma$
above, and these occur with coefficient~$1$. Since the parts of $\sigma$ have
size
at most~$d$, this means that the parts of $(N-d)^d - \nu$ have size at most~$d$,
and this is the partition~$\lambda$ from the lemma statement. Tracing backwards,
we see that $\nu$ is $(M-d)^d + \lambda'$, and thus the expression of the
correspondence consists of the classes $[\lambda] \otimes [d^d - \lambda']$, as
in the statement.
\end{IEEEproof}

\begin{IEEEproof}[Proof of Theorem~\ref{t:symmetric}]
It will be sufficient to prove that the product of the classes from
Lemma~\ref{l:containment-class} over all pairs
$i \neq j$ results in a positive multiple of some Schubert class. Moreover, by
Theorem~\ref{t:schubert-non-negativity}, it is sufficient to find one
combination of terms from each incidence class whose product is non-zero. We
shall exhibit such a combination in two separate cases.

\begin{figure}
\begin{centering}
\includegraphics{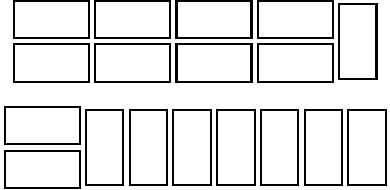}
\par\end{centering}
\caption{Schematic diagram of the partitions whose product gives a non-zero
coefficient in the proof of Theorem~\ref{t:symmetric} for the case when $K$ and
$d$ are even. On top are the partitions for the receiver's Grassmannian and on
the bottom is the transmitter's Grassmannian, when the index is odd. Each block
represents a single copy of $d^{d/2}$ or $(d/2)^d$, and the arrangement shows a
partition with non-zero coefficient in their product.}
\label{f:SchematicPartitions}
\end{figure}
First, we suppose that $K$ is odd. From the factor for each incidence
correspondence, we choose the term based on the cyclic difference $(i-j) \bmod K
\in \{1, \ldots, K-1\}$, where $j$ is the index of the receiver and $i$ is
the index of the transmitter. In particular, when this modular difference is between
$1$ and $(K-1)/2$ inclusive, we choose the term where the transmitter partition
is~$d^d$. We write $a^b$ to denote the partition consisting of $b$ parts of size~$a$. By
Lemma~\ref{l:containment-class}, the term has the empty partition~$0$ for the
receiver's Grassmannian. When this modular difference is at least $(K+1)/2$, we
choose the term where the receiver partition is $d^d$ and the transmitter
partition is~$0$. Thus, for each Grassmannian, we have the product of $(K-1)/2$
copies of $d^d$. We have assumed that $N-d \geq d(K-1)/2$, so this product
contains at least one copy the tuple with $(d(K-1)/2)^d$ in each spot.

Suppose now that $K$ is even and $d$ is also even.
We choose the term of each incidence relation as follows (schematically depicted
in Fig.~\ref{f:SchematicPartitions}). The transmitter's Grassmannian
has partition $(d/2)^d$, with the exception that when the transmitter's index $j$ is
even and the transmitter's index~$i$ is equal to $j+1$ or $j+2$, modulo $K$, the
transmitter has $d^{d/2}$. In either case, the receiver's partitions are the
same.
For each receiver, we have the product of $K-2$ copies of $d^{d/2}$ and
one $(d/2)^d$. The product of $d^{d/2}$ with itself has a non-zero coefficient
in front of $d^d$. Then, the product of $(K-2)/2$ copies of $d^d$ with $(d/2)^d$
has a non-zero coefficient in front of $(d(K-1)/2)^d$, using our assumption
that $N-d \geq d(k-1)/2$. At a transmitter with odd index, we are evaluating the
product of $K-1$ copies of $(d/2)^{d}$, yielding a non-zero coefficient in front
of $(d(K-1)/2)^d$. At an even index, it is similar except two of these copies
are replaced by $d^{d/2}$, which themselves multiply to $d^d$, and we get the
same result.

Finally, when $K$ is even and $d$ is odd, the proof is similar, except that
since $d/2$ is not an integer, we have to round it up or down each time it is
used. In particular, for the factor corresponding to the incidence relation
between the $i$th receiver and $j$th transmitter, we use the same partitions above
except that we round down $d/2$ in the transmitter's partition when $i-j$ is even and
round up when $i-j$ is odd. Of course, this causes $d/2$ in the receiver's
partition to round in the opposite direction. Overall, for each Grassmannian
we have rounded up half the time and down half the time and the product works out
as above.
\end{IEEEproof}

The assumption of a symmetric alignment problem was crucial in being able to
choose rectangular Schubert classes in the proof of
Theorem~\ref{t:symmetric}. The difficulty of generalizing this proof to
non-symmetric problems is to find Schubert cycles whose product is
provably positive, and unlike the symmetric case, rectangular partitions may not
suffice in general.

\section{Computing feasible strategies}
\label{sec:computing}
In this section, we discuss the computation of
strategies from particular channel matrices.
In the case of $K=3$, the proofs of sufficiency in
Propositions~\ref{p:k-3-symmetric} and~\ref{p:k-3-critical} are effective, in
that the feasible strategies can be computed via eigenvectors or the kernel of a
matrix, respectively.
For $K > 3$, the feasibility problem is not reduced
to linear algebra, and the method of Theorem~\ref{t:symmetric} is not constructive,
but we can still solve the interference alignment problem using general numerical methods
for polynomial equations.

\begin{table}[tb]
\caption{Number of solutions to symmetric alignment problem}
\label{tbl:numbers-solutions}
\begin{centering}\begin{tabular}{r|rrr}
& \multicolumn{3}{c}{$d$} \\
$K$ & 1 & 2 & 3 \\
\hline
3 & 2 & 6 & 20 \\
4 & - & 3700 & - \\
5 & 216 & 388,407,960 &  \\
6 & - & & - \\
7 & 1,975,560 & & \\
8 & - & & - \\
9 & 2,355,206,975,800\\
\end{tabular}\par\end{centering}
The table shows the number of solutions to the symmetric alignment problem when
$N = M = d(K+1)/2$ and either $d$ or $K+1$ is even. Missing values could not be
computed.
\end{table}

The method used to prove Theorem~\ref{t:symmetric} was to establish a lower
bound on the generic number of solutions using Schubert calculus. It was
sufficient to find one product of classes which was positive. However,
by computing all terms in the product of the incidence relations, we
can compute the number of solutions for a general system in small cases:

\begin{proposition}\label{p:numbers-solutions}
Suppose that either $d$ is even or $K$ is odd. Then the number of solutions to
the symmetric alignment problem is as given in
Table~\ref{tbl:numbers-solutions}.
\end{proposition}

Using very different methods, \cite[Sec.~IV]{GS13} counted the approximate
number of solutions to some interference alignment problems. In the common
cases, the two methods agree up to their stated margins of error. In particular,
we confirm that for $M=N=5$, $K=4$, and $d=2$, there are 3700 solutions, which
they could only claim with high confidence. In addition, the first two
values in Table~\ref{tbl:numbers-solutions} for $K=3$ were computed
in~\cite{SUH10} using Bernstein's Theorem.

The solution counts given in Proposition~\ref{p:numbers-solutions} are relevant
for computing solutions in two different ways. First, a large number of
solutions indicate the difficulties in enumerating all solutions or in using an
iterative algorithm.

Second, the number of solutions also measures the algebraic
complexity of finding a solution, since it is also the degree of the field
extension
of a solution for random rational parameters. To illustrate this
principle, consider the case of
Proposition~\ref{p:k-3-symmetric}, which showed that the solution to the interference
alignment problem with $d=3$ and $N=M=2d$ can be found by finding $d$
eigenvectors of a $2d \times 2d$ matrix. Algebraically, finding an eigenvalue
and eigenvector of a generic $2d \times 2d$ matrix with rational entries
requires solving an irreducible polynomial of degree $2d$, and thus the solution
lies in a degree~$2d$ extension of the rationals. Finding a second eigenvector
requires an extension of degree $2d-1$, and so on, so that the solution lies in
a degree~$(2d)! / d!$ extension. A somewhat more refined analysis would show
that it lies in a smaller field of degree $\binom{2d}{d}$, but in any case the
prime divisors of the degree are less than $2d$, and likewise the polynomials
that had to be factored had degree less than~$2d$. The structure of the field
extension and its degree reflect the structure of the construction of the
solution.

In contrast, Proposition~\ref{p:numbers-solutions} shows us that when, for
example, $K=4$, $d=2$ and $N= M =5$, for sufficiently general rational data, the
solution lies in an extension of the rationals of degree $3700 = 2^2 \cdot 5^2
\cdot 37$. Thus, an exact solution must (explicitly or implicitly) entail finding a root of a polynomial of degree~$37$, or, worse, some
multiple of~$37$. It seems unlikely that there is any natural construction
for such a polynomial.

Thus, when $K > 3$, we turn to numerical methods, such as
PHCpack~\cite{phcpack},
Bertini~\cite{bertini}, and HOM4PS~\cite{hom4ps}, which use homotopy
continuation methods to solve arbitrary algebraic equations.
Specialized homotopy methods for Schubert problems in a Grassmannian were
introduced in~\cite{HuberSottileSturmfels} and~\cite{SottileVakilVerschelde}.
However, to our knowledge there currently is no implementation of these ideas
for Schubert problems in a \emph{product} of Grassmannians, such as our strategy space.

To represent the
unknown vector spaces in the alignment problem, we could use Pl\"ucker
coordinates.
However, computationally, the Pl\"ucker coordinates are inefficient and lead to
systems with more equations than unknowns, which are more difficult for the
numerical solvers. Instead, we choose special coordinates for which our system
is square, following~\cite{HauensteinHeinSottile}.

Under our genericity assumptions, we can assume that each $U_i$ is generated by
the rows of a $d_i \times N_i$ matrix with an identity in the leftmost position:
\begin{equation*}
\begin{pmatrix}
1 &  \cdots & 0 & u_{i,1,1} & \cdots & u_{i,1,N-d} \\
\vdots &  \ddots   & \vdots &  \vdots  & & \vdots \\
0 & \cdots & 1 & u_{i,d,N-d} & \cdots & u_{i,d,N-d}
\end{pmatrix},
\end{equation*}
and similarly for~$\overline V_j$. The orthogonality conditions between $V_i$ and $U_j$
can be expressed as $d_i d_j$ bilinear equations in these variables. Note that
in this representation, the expected dimension ($t_{\{1, \ldots, K\}}$
in Theorem~\ref{t:upper-bound}) is the difference between the numbers of
variables and the number of equations.

\begin{example}
We consider the symmetric alignment problem with $K=5$, $M=N=3$, and $d=1$.
Using Macaulay2~\cite{macaulay2}, we chose the $20 = 5 \cdot (5-1)$ channel
matrices~$H_{ij}$ as random $3 \times 3$ matrices with rational entries. Each
of the $1$-dimensional vector spaces is given as the span of a vector $(1, *,
*)$, and thus there were $10 \cdot 2$ coordinates. In addition, each of the
channel matrices gives one condition, so there were also $20$ equations of the
form:
\begin{equation*}
\begin{pmatrix} 1 & v_{i,1} & v_{i,2} \end{pmatrix}
H_{ij}
\begin{pmatrix} 1 & u_{j,1} & u_{j,2} \end{pmatrix}
\qquad\mbox{for } i \neq j
\end{equation*}
We used PHCpack~\cite{phcpack} to
solve this system and obtained $216$ solutions in about a minute and a half on a
laptop, thus confirming the value in
Table~\ref{tbl:numbers-solutions}.
\end{example}

\section{Conclusion}
\label{sec:conclusion}
This paper makes progress in understanding the role of diversity in aligning
interference by focusing on spatial diversity due to multiple antennas. While
the performance benefits under the model we have considered are limited, our
larger goal has been the introduction of new methods of analysis.
The approach we used in~\cite{BCT11} and~\cite{BCT11a} to study vector space strategies for interference alignment has in a short time led to progress in a variety of other channel models and scenarios. For example, under different models in which interaction is allowed between transmitters and receivers, both Annapureddy et al~\cite{AEV11} and Geng et al~\cite{geng2013interactive} use algebraic techniques to show that substantially better performance may be possible.

\appendix[Proof of Lemma~\ref{l:AfullRank}]
\label{sec:fullRankProof}
In order to prove that $\bA_r$ has full rank for generic channel matrices, it is
sufficient to prove that it does so for one particular set of matrices. This is
because $\bA_r$ having full rank is equivalent to at least one of its maximal
minors being non-zero. If this is true for one specialization of the channel
matrices, then it will be true for a dense open set.
We specialize to the matrices
\begin{equation*}
B := \bH{i_{j}}{i_{j+1}}
= \begin{pmatrix} I_{M} \\ \mathbf 0 \end{pmatrix}
\end{equation*}
and
\begin{equation*} C := \bH{i_j}{i_{j+2}}
= \begin{pmatrix} \mathbf 0 \\ I_{M} \end{pmatrix},
\end{equation*}
where $I_M$ denotes the $M\times M$ identity matrix and the $\mathbf 0$ denotes
a block of $0$s of size $(N-M) \times M$.
Note that by our assumptions on the sequence of indices $i_1, \ldots, i_{r+2}$,
we have made only one
assignment for each matrix in (\ref{e:big-matrix}).

We will prove that, with these specializations, the block matrix~$\bA_r$ from
the statement has full rank by
simultaneous induction on~$r$, $N$, and~$M$. If $r=0$, then $\bA_r$ is a $0
\times M$ matrix, which trivially has full rank. If $N \geq 2M$, then every row
vector is a unit vector and all such unit vectors appear in some row, so the
matrix has full rank.

Now we suppose that $N < 2M$.
We permute the rows and columns of $\bA_n$ as follows. Note that the rows
of~$\bA_n$ are arranged into $r$~blocks of $N$ rows each. We extract the first
block in its entirety, followed by the last $N-M$ rows of each of the subsequent
$r-1$ blocks. We leave the remaining rows in their induced order, and put
extracted rows after them, also in their induced order.
We also permute the columns, which are arranged into $r+1$
blocks of size~$M$. We take the first block of $M$~columns, followed
by the last $N-M$ columns of each of the other $r$ column blocks, and place
these to the right of all the other columns.

If we divide $B$ and~$C$ into blocks by separating off the last $N-M$ rows and
columns of each, then we get
\begin{equation*}
B = \begin{pmatrix} \tilde B & B' \\ \mathbf 0 & \mathbf 0 \end{pmatrix}
\quad
C = \begin{pmatrix} \tilde C & \mathbf 0 \\ \mathbf 0 & I_{N-M} \end{pmatrix}\,,
\end{equation*}
where
\begin{equation*}B' = \begin{pmatrix} \mathbf 0 \\ I_{N-M} \end{pmatrix}
\quad
\tilde B = \begin{pmatrix} I_{2M-N} \\ \mathbf 0 \end{pmatrix}
\quad
\tilde C = \begin{pmatrix} \mathbf 0 \\ I_{2M-N} \end{pmatrix}\,.
\end{equation*}
Therefore, the rearranged matrix has the form
\begin{equation*}
\begin{pmatrix}
\tilde B&\tilde C&        &        &        &B'  \\
        &\ddots  &\ddots  &        &        &        & \ddots \\
        &        &\tilde B&\tilde C&        &        &       & B' \\
\tilde C&        &        &        & I_{M}  &        &  \\
        &        &        &        &\mathbf0&I_{N-M} &  \\
        &        &        &        &        &\mathbf0&I_{N-M}& \\
        &        &        &        &        &        &\ddots & \ddots \\
        &        &        &        &        &        &       &\mathbf 0 & I_{N-M} \\
\end{pmatrix}
\end{equation*}
In the lower right, we have a identity matrix of size $M+r(N-M)$.
We can use this identity matrix, together with elementary column operations to
clear the~$\tilde C$ on the left, and with elementary row operations, the~$B'$s
in the upper right.
The transformed matrix is in block diagonal form, with an identity matrix in the
lower right. In the upper left, the copies of $\tilde B$ and~$\tilde C$ form our
specialized version of $\bA_{r-1}$ with parameters $M$ and~$N$ each decreased by
$N-M$. By the inductive hypothesis, the latter matrix has full rank.
Thus, $\bA_r$ is equivalent to a block diagonal matrix,
where one block has full rank, and the other block is the identity, so $\bA_r$
has full rank.

\section*{Acknowledgment}
We thank Bernd Sturmfels for insightful discussions and the authors of
\cite{RGL11} for sharing their related manuscript with us before it was
publicly available.
We also wish to acknowledge Frank Sottile for suggesting the method of proof in
Theorem~\ref{t:symmetric}.

\bibliographystyle{ieeetr}
\bibliography{BIB}

\end{document}